\documentclass[10pt,twocolumn]{article}
\pdfoutput=1 

\usepackage[round]{natbib}
\usepackage{subcaption}
\usepackage{float}
\usepackage{graphicx} %to import figures
\graphicspath{images/} %where to find figures, make sure to convert jpeg as png for easy importing
\usepackage{url}

\usepackage{breakurl}
\usepackage[breaklinks]{hyperref}
\usepackage{amstext}
\usepackage{tabu}
\usepackage{moreverb}
\immediate\write18{texcount -inc -incbib 
-sum research_notes.tex > /tmp/wordcount.tex}

\usepackage{titlesec}
\titleformat{\section}
  {\normalfont\fontsize{12}{15}\bfseries}{\thesection}{1em}{}

\usepackage[utf8]{inputenc} % allow utf-8 input
\usepackage[T1]{fontenc}    % use 8-bit T1 fonts
\usepackage{booktabs}       % professional-quality tables
\usepackage{amsfonts}       % blackboard math symbols
\usepackage{nicefrac}       % compact symbols for 1/2, etc.
\usepackage{microtype}      % microtypography
\usepackage{lipsum}

\title{Airbnb's disruption of the housing structure in London}

\author
{Zahratu Shabrina,$^{1}$ Elsa Arcaute,$^{1}$ Michael Batty $^{1}$\\
\\
\normalsize{$^{1}$Centre for Advanced Spatial Analysis (CASA), University College London, UK}\\
\normalsize{$^\ast$To whom correspondence should be addressed; E-mail:  zahratu.shabrina.15@ucl.ac.uk.}\\
}

\begin{document}
\maketitle
\begin{abstract}
\noindent
This paper explores Airbnb, a peer-to-peer platform for short-term rental of housing accommodation, examining the geographical pattern of those establishments using data from London. Our purpose is to analyse whether or not the diversity of dwelling types correlate with the distribution of listings. We use a measure of spread based on entropy to indicate the diversity of dwelling types and look at its relationship with the distribution of Airbnb establishments, as well as the type of home ownership using correlation analysis. It is important to note that our study only considers domestic building types, and excludes any information on the diversity of land uses. Two important findings emerge from our analysis. Firstly, the spatial location of Airbnb rentals is negatively correlated with the diversity of dwelling types, and positively correlated with a single dwelling type, which corresponds in general to purpose built flats, conversions and flats in commercial buildings. Secondly, Airbnb is associated with areas that have a high proportion of privately rented properties, detracting more than 1.4\% of the housing supply into short-term rentals. Such a phenomenon can reach up to 20\% in some neighbourhoods, further exacerbating the process of gentrification. Finally, we discuss the implications of these findings as instruments to inform policies associated with the 'sharing' economy in relation to the disruption of the housing structure.\\

\end{abstract}

\section{Introduction}
London has the status of a city with extreme inequalities \citep{hamnett2004unequal,watt2016london} composed of rapidly rising rents and house prices, the separation of social classes and complex multifaceted gentrification \citep{glass1964aspects,davidson2012class}. There have been dramatic economic and social changes in the past five decades, affecting both its social structure and the built environment. One of the dominant features of these effects has been in London's housing market. \cite{hamnett2004unequal} argues that its housing market is unequal due to two main reasons, (1) the nature of the relationship between  housing market and labour market locations coupled with income, and (2) changes in London's occupational and earning structure. The rise of high-income professional and managerial workers over time has helped polarise London's housing structure with the emergence of very high priced property especially in inner-city areas \citep{hamnett2004unequal, watt2016london}. London experiences a strong drive at the top end of the market which trickles down to all sectors, further pushing up  house rents and prices generally. As a world city, London is not alone in this, as other global cities such as New York, Hong Kong, and Paris are experiencing similar trends. In these cities, it has become increasingly challenging to obtain home ownership for average or below-average income residents. There are also challenges in finding long-term rental properties that are not overvalued. 

Recently, there has been an explosion of short-term rental platforms such as HomeSwap, HomeAway and Airbnb. These are considered as part of the platform ('sharing') economy, which refers to the open access approach of utilising assets and services \citep{richardson2015performing,rifkin2000age}, mostly for profit \citep{belk2014sharing}, and generally mediated by Internet technology \citep{schor2016debating, agyeman2013sharing}. These hospitality platforms offer two main benefits to those who participate: economic incentives for hosts from renting their underutilised rooms or homes, and choices of short-term accommodation with extra amenities often not available at hotels for guests. The biggest short-term rental platform remains Airbnb, a San Francisco based private technology company offering accommodation in a marketplace that connects hosts (with entire apartments, private or shared listings) and guests. As a mediator, Airbnb acquires a designated fee for each successful booking. According to their website, in 2018 there were nearly 5 million Airbnb listings in 81,000 cities located in more than 191 countries worldwide. Based on a study using semi-structured interviews with Airbnb hosts who use the platform, one of the most repeated extrinsic motivations for renting through this peer-to-peer market is to supplement rent or mortgages \citep{ikkala2015monetizing, jefferson2014airbnb}. This is also true in the case of London, in which sharing accommodation or long-term subletting is a very common practice to lessen the burden of the cost of renting or owning a house. This might be one of the reasons why Airbnb is very popular in cities where the standard of living is relatively high, mainly because it provides a channel for extra earnings. According to the dataset from Inside Airbnb (http://insideairbnb.com) which has collected publicly available data from the Airbnb website, the top three most popular Airbnb locations are London, with more than 69,0000 listings in 2018, followed by Paris and New York respectively, all cities where the living costs are above average.   

In London, section 25 of the Greater London Council (General Powers) Act 1973 was previously used to restrict the use of residential premises as temporary sleeping accommodation, protecting London housing supply and benefiting permanent residents by preventing the conversion of family homes into short-term lets. But on the 26th May 2015, London implemented Section 44 of the Deregulation Act \citep{deregulationact} allowing Londoners to rent out properties for up to 90 days without planning permission, providing there is no change of use (for which planning permission would be required) \citep{deregulationactchange}. Before the Deregulation Act, unlawful short-term rentals could cost homeowners a fine of up to £20,000, but this would not affect subletting by renters who are mostly restricted by the tenancy agreement. This change in regulation might give way to short-term rental platforms widely spread across London.

This paper aims to frame the nature of the housing crisis in London, a global city that has undergone extreme distortions in the housing market \citep{edwards2016housing, watt2016london} and has widely adopted digitally mediated short-term rental platforms \citep{guttentag2015airbnb, gutierrez2017eruption, volgger2018adopts, guttentag2017assessing} which are being identified as invading the long-term rental market in various cities globally \citep{lee2016airbnb, horn2017home}. Specifically, we explore how the geographical location of Airbnb is related to dwelling types, and London's housing structure in terms of the type of housing ownership. It is important to emphasise that we only focus here on domestic building types, thus excluding commercial buildings and other functions. It is beyond the scope of this paper because Airbnb is used mainly for accommodation and not for other functions. The next section provides the background to our study, followed by the data and methodology used to analyse the problems we identify. We then present the results and close with concluding remarks.

\section{The complex problems of the London housing market}
Even before Airbnb was introduced, London has undergone a long history of complex housing problems. As one of the most expensive cities to live in the world, London is now experiencing an acute housing shortage and affordability crisis. \cite{hilber2015uk} found that for similar flats, the rental price-per-square meter in London is the second highest in the world, below Monaco, and followed only by Hong Kong, New York and Paris. The roots of London's housing problems can be linked to two causes. The first is that despite recent efforts of new house construction over the last 10 years, there has been a prolonged decline in new houses constructed since the 1970s, leading to a continuous housing shortfall, thus creating severe supply constraints \citep{londonhousing, watt2016london}. Another driver of this supply constraint is the scarcity of developable land in London \citep{hilber2010impacts}. The second problem involves regulatory constraints as well as housing-related policies. A study by \cite{hilber2010impacts} suggests that regulatory restrictiveness raises house prices more than it otherwise would. Also, some policies formulated with an intention to help first time buyers, such as 'Help to Buy' schemes that help with down payments, eventually pushes up house prices further due to increase in demand that is followed by unresponsive supply \citep{hilber2013help}.

Nowadays, the rapid growth of short-term rentals such as Airbnb is feared to have worsened the housing crisis and affordability in many cities worldwide. Past research has examined Airbnb from the important perspective of exploring the effects of short term rentals on the housing market and gentrification \citep{lee2016airbnb, horn2017home, wachsmuth2017airbnb}. Although some have argued that Airbnb significantly disrupts the hotel industry with an increase in Airbnb listings resulting in a decrease in hotel room revenue \citep{guttentag2015airbnb, zervas2016rise}, it is important to note that Airbnb listings expand holiday rentals in residential areas as well. That is why it is important to look at the possible negative long and short-term effects on the housing market. Based on the analysis by \cite{wachsmuth2017airbnb} who examine the relationship between gentrification and short-term rentals, Airbnb forces a new kind of rent gap that is fused by the new revenue flow from the platform. This is happening mainly in areas that are currently gentrifying or that are post-gentrified neighbourhoods. \cite{horn2017home} revealed that a one standard deviation increase in Airbnb listings results in an 0.4\% increase in housing rent in Boston due to the decrease in the supply of units available for prospective long-term residents. This finding is also supported by \cite{lee2016airbnb} who studied Airbnb and the housing crisis in Los Angeles. Moreover, there are many misuses through short-term rentals in the housing market where hosts list more than one offer, for example in Berlin where as many as 0.30\% (5,555 properties) of the total housing stock are being misused as Airbnb listings \citep{schafer2016misuse}.

This paper assesses the linkage between the two topics: housing and the short term rental phenomenon in London. This is especially important in a city such as London where there is a rising trend of owning property as an investment. London is the centre of a speculative market, with many overseas investors as well as buy-to-let landlords buying properties without any intention of living in the properties themselves or even renting them out \citep{dorling2014all}. This is especially true in inner-city areas, where multi-occupancy housing for rents is very common. Additionally, as short term rentals become a more attractive and viable option, private companies specialising in managing these properties have emerged. They are referred to as 'Airbnb management', a business model that manages Airbnb rentals (cleaning, keys management, website update, etc) in exchange for a share of the profit, eliminating the need for owners to be present to manage their own properties. These short-term rental platforms have become a bigger business model for both micro-entrepreneurs and businesses alike.

\section{The geographical pattern of Airbnb establishments in London}
Here we use several datasets to analyse and visualise the relationship between Airbnb and housing in London, in association with dwelling type diversity. We use the \textit{Airbnb data} from Inside Airbnb (http://insideairbnb.com) which has information on the rental outlets that were available for booking in June 2018. There is a total of more than 69,000 observations (Airbnb rentals) in London, from which 50,000 (72\%) have at least one review, which can be used as a proxy for active listings on the platform. Listings refer to properties (entire properties, private rooms or shared rooms) that are being advertised on the Airbnb platform. Each listing has attributes such as estimated geographical location, price per-night, number of beds/bedrooms, number of reviews, and so on. The data shows that 55\% are listings of entire property, 44\% are private rooms and only 1\% correspond to shared rooms.
We combine this Airbnb data with \textit{dwelling types data from the UK Census 2011}. The dwelling types in the UK are divided into three main categories, whole house or bungalow (further divided into detached, semi-detached and terraced); flats, maisonettes or apartments (further divided into purpose-built, conversions or flats in commercial buildings); and mobile or temporary structures. We also use the housing supply data from \textit{Valuation Office Agency (VOA)}. The Valuation Office Agency data consists of the number of houses in each London LSOA (Lower Super Output Area) and this makes up the total building stock available for London dwellings. Lastly, we use the \textit{LSOA type of home ownership data} from the UK Census 2011 which contains a summary of areas with proportion of the type of home ownership.

\begin{figure}
        \centering
            \includegraphics[width = 0.23\textwidth]{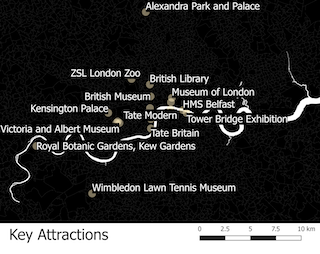}
            \includegraphics[width = 0.23\textwidth]{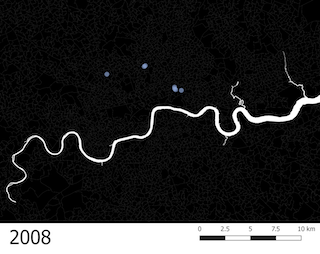}
            \includegraphics[width = 0.23\textwidth]{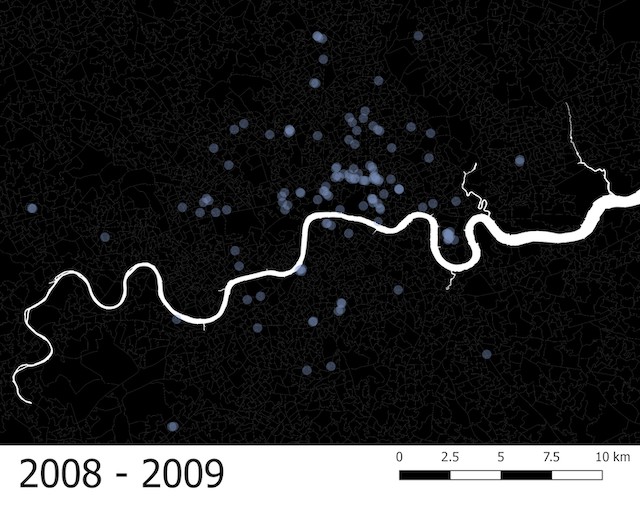}
            \includegraphics[width = 0.23\textwidth]{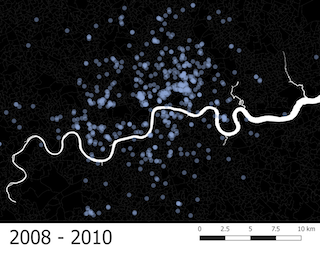}
            \includegraphics[width = 0.23\textwidth]{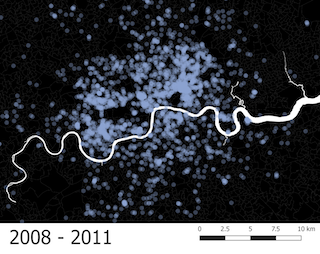}
            \includegraphics[width = 0.23\textwidth]{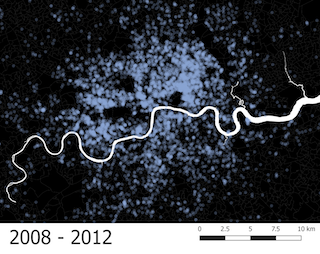}
            \includegraphics[width = 0.23\textwidth]{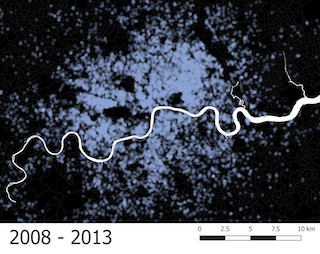}
            \includegraphics[width = 0.23\textwidth]{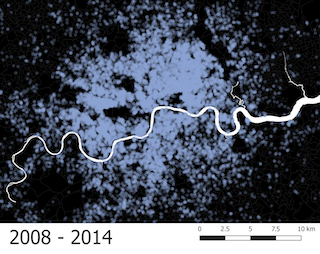}
            \includegraphics[width = 0.23\textwidth]{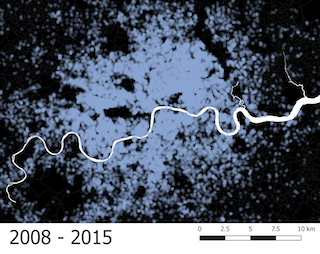}
            \includegraphics[width = 0.23\textwidth]{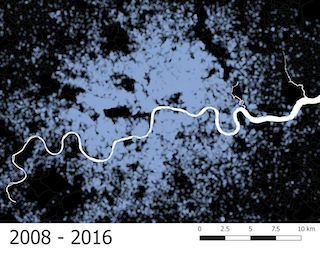}
            \includegraphics[width = 0.23\textwidth]{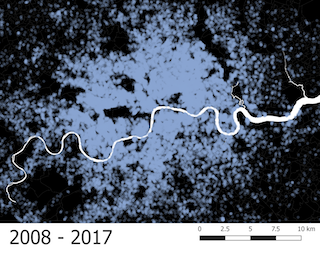}
            \includegraphics[width = 0.23\textwidth]{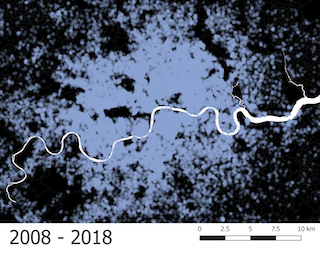}
           
    \caption{Airbnb listings accessible for bookings on June 2018 according to the time of first listing} 
    \label{fig:temporal}
\end{figure}

We start by examining how Airbnb listings have survived over the years. We divide our data into subsets of longitudinal data based on the year the host first listed their Airbnb listings. Figure \ref{fig:temporal} provides a series of cumulative temporal snapshots of the current Airbnb listings, i.e. it shows the set of listings available (from the currently existing ones) for booking at the specific time interval in Central London. The far left of the Figure \ref{fig:temporal} shows the key attractions in London based on Visit London 2015 Data, that showcases the concentration of tourist attractions around the River Thames. Figure \ref{fig:temporal} shows that very few Airbnb listings currently available have been hosted since 2008, and they are all located north of the river (there is the total of eight listings). The distribution of London tourist attractions is quite concentrated in the central areas north of the river, and this aligns with the distribution of older Airbnb establishments. The previous study by \cite{li2016pros} finds that the turnover of establishments on the Airbnb platform is relatively high, where 49\% of hosts exited the market after registering (based on 18 months of observations of Airbnb in 2012 to 2013). Also, non-professional hosts (hosts with one listing instead of multiple listings) are 13.6\% more likely to exit the market \citep{li2016pros}. Unfortunately, our data cannot provide the overall evolution and growth of Airbnb because it does not contain the listings that have disappeared from year to year. But even disregarding these disappearing listings, our visualisation showcases how Airbnb listings increase cumulatively over the years to the point in 2018 where London is now populated rather heavily with Airbnb.

\begin{figure}
    \centering
    %\begin{subfigure}{0.49\textwidth}
            \includegraphics[width = 0.49\textwidth]{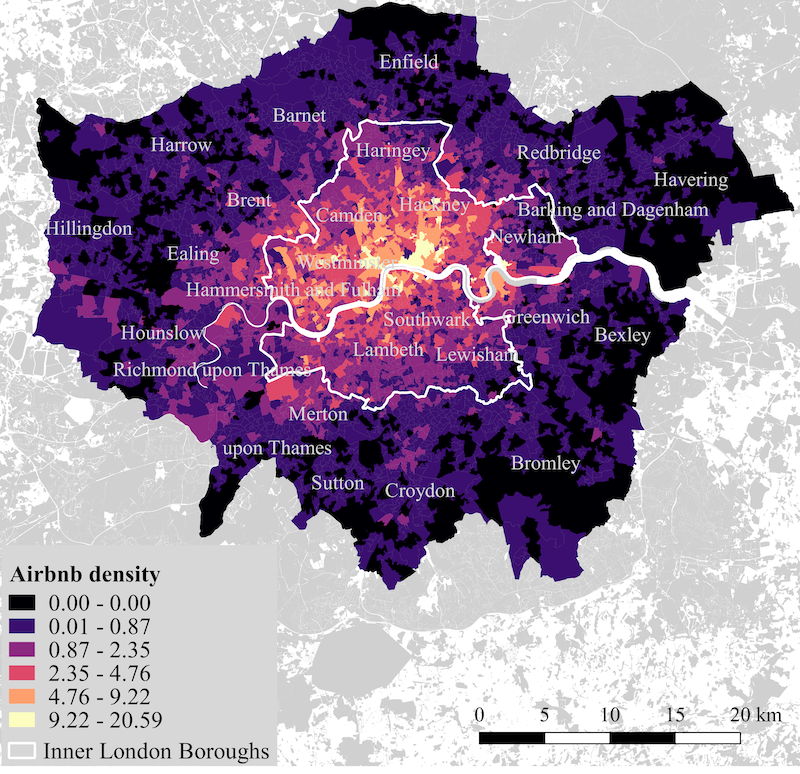}
    \caption{Airbnb density in London compared to the housing supply in London LSOAs}
    \label{fig:airbnb_dist}
\end{figure}

Figure \ref{fig:airbnb_dist} further aggregates the Airbnb data into LSOAs which typically have a population of around 1500 in London. We also differentiate between inner and outer boroughs by creating an outline of inner city areas (the white boundary on the maps) and outer areas. This outline follows the classification proposed by the London Council for Inner and Outer Borough area based on Ordnance Survey (OS) data. London is quite monocentric in terms of its activities and employment centres \citep{buck2013working}, so we would wish to see if this is also the case for the Airbnb distribution. We can see that although Airbnb intensity tends to increase in the city centre as has been confirmed in previous studies by \cite{dredge2015collaborative} and \cite{ gutierrez2017eruption}, the Airbnb phenomenon clearly extends beyond the central areas. We have examined the Airbnb distribution in all 4835 LSOAs and found that 82\% of the areas (3982 LSOAs) have at least one Airbnb listing. This also aligns with the findings by \cite{coles2017airbnb} suggesting that although centrality is the main driver in Airbnb locations, Airbnb listings have become geographically more dispersed than many other comparable establishments such as hotels, guest houses and hostels.  

Based on the Valuation Office Agency data, there are approximately 3.4 million dwellings in London, and compared to this number, houses advertised through the platform are only 1.4\% of the overall housing supply. This number seems small and somewhat insignificant, but because Airbnb is very concentrated in several areas, some regions are more affected compared to others. Figure \ref{fig:airbnb_dist} shows the density of Airbnb listings calculated by comparing the number of Airbnb establishments with the housing supply. The limitation is that we assume that any type of listing (regardless of whether they are entire property, private or shared listings) is equal to one unit of housing converted to short-term rentals. The dark colours indicate low density while the lighter colours indicate high density of Airbnb establishments. When compared with the overall housing supply, the highest Airbnb density is found in LSOAs located in Hackney (up to 21\%), Tower Hamlets (up to 19\%), Westminster (up to 18\%), Camden (up to 14\%) and Southwark (up to 11\%) respectively. When looking at this phenomenon locally, these numbers are quite significant considering that these houses detract from the long term rental market.  

% There is an LSOA in Hackney where more than 20\% of the houses have been advertised through the platform. The highest Airbnb density also found in LSOAs is located in Tower Hamlets, Westminster, Camden and Southwark where more than 10\% of the housing supply is being advertised as Airbnb listings. This number is very significant considering that these houses detract from the long term rental market.

\cite{coles2017airbnb} found that Airbnb is most profitable compared to long-term rentals in middle-income neighbourhoods. Areas such as Tower Hamlets and Hackney are among those with highest Airbnb listings, and according to \cite{Savills2014} these two middle-income areas have been experiencing the highest influx of affluent newcomers and creative workers over recent years \citep{Savills2014}. From Figure \ref{fig:airbnb_dist}, we can also see the areas in which 9\% to 20\% of the housing has been rented as entire property listings via the Airbnb platform, which is mainly in well-developed and gentrified neighbourhoods. These areas are most prone to the negative effects of Airbnb especially with respect to the rise in housing prices.  

\section{Entropy of dwelling types in London}
On further examination, we analyse if the proportion of certain dwelling types can be associated with the distribution of Airbnb listings in London. We use accommodation data from the 2011 UK Population Census that refers to the types defined by the use of the dwellings rather than the typology of the buildings. Table \ref{tab:dwellingtype} shows the structure of dwelling types in the Greater London Area based on their count and overall proportion. Almost 48\% of dwellings in London are whole houses or bungalows and the rest, 52\%, are flats or apartment types of dwelling (dominated by purpose-built flats or maisonettes that make up the largest proportion of almost 38\% of London dwellings).

\begin{table}
    \centering
    \begin{tabular} {| l  l   l  |} 
 \hline
\textbf{Dwelling Type} & \textbf{Count} & \textbf{Pct} \\ [0.5ex] 
 \hline\hline
\textbf{Whole house or bungalow} & & \\
\hline
Detached &    211232 & 6.24\% \\ 
\hline
Semi-detached & 629607 & 18.59\%\\
 \hline
Terraced &    776821 & 22.93\% 
\\
\hline
\textbf{Flat, maisonette or apartment} & & \\
\hline
 Purpose-built block of flats &1274526 & 37.63\%
 \\
 \hline
Conversion  &429456 &     12.68\%
\\
\hline
In a commercial building & 62795 &     1.85\%
\\
\hline
\textbf{Mobile/temporary structure} &2818 &     0.08\%
\\\hline
\textbf{Total} & \textbf{3387255} & \textbf{100\%}
\\\hline

\end{tabular}
    \caption{Proportion of dwelling types in the Greater London Area (Source: UK Census 2011)}
    \label{tab:dwellingtype}
\end{table}

Using the data, we calculate the entropy for the distribution of dwelling types in  each London LSOA. Entropy is a prominent measure which lets us study the complexity of cities in terms of their functional mix and heterogeneity \citep{batty2014entropy, wilson2013entropy}. It is a measure of spatial variance that is at a maximum when the distribution of types is uniform or completely heterogeneous and is at a minimum when the distribution is peaked around a single type thus representing extreme homogeneity. The concept of entropy has been used in geography for different research purposes ranging from measuring urban land use change and composition \citep{zhao2004study, tan2003laws}, to describing the geographic concentration of economic activity \citep{garrison1973entropy}, and urban spatial interaction modelling \citep{jat2008monitoring} but here we use it in its statistical form. The measure of entropy used is originally due to Shannon where information varies inversely with the probability of occurrence. The probability of dwelling types $ P_i $ can thus be calculated as follow:

\begin{equation} \label{eq:1}
    {P_i = \frac{A_i} {\sum\limits_{i}^{K}{A_i}}}
\end{equation} where \textit{$ A_i $} and \textit{K} represent the number of dwelling types \textit{i} and the total number of observed dwelling types respectively. Entropy denoted as \textit{H} is calculated as:
\begin{equation} \label{eq:2}
H = - \sum\limits_{i}^{K}{P_i * \ln P_i}
\end{equation} while its maximum $(H_{max})$ can be computed as:
\begin{equation} \label{eq:3}
 H_{max} = \ln K
\end{equation}

\begin{figure}
        \centering
            \includegraphics[width = 0.50\textwidth]{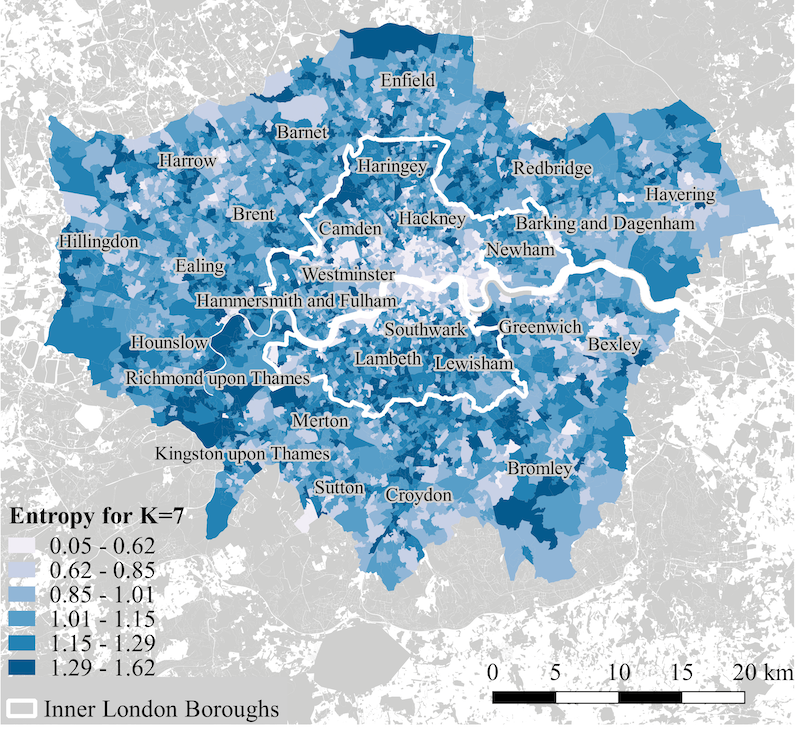}
    \caption{Entropy of dwelling types in London defined by the methodology outlined in the text. The darker colours indicate high diversity of dwelling types while lighter colours indicate less diversity and more homogeneity}
    \label{fig:entropy}
\end{figure}

Entropy is a measure of heterogeneity with higher values indicating a higher number of dwelling types in an area (with more diversity or a greater mix). Figure \ref{fig:entropy} visualises the entropies for all seven dwelling types calculated using Equation (\ref{eq:1}) and (\ref{eq:2}). The darker colour blue indicates higher entropy values, meaning that those areas have a greater diversity of dwelling types. Maximum entropy is achieved when all types are equally likely thus presenting maximum diversity \citep{atkinson2014scale}. Maximum entropy is defined using Equation (\ref{eq:3}) and therefore the maximum value for our case is approximately 1.94 and the minimum entropy is 0. Minimum entropy 0 indicates an area dominated by a single dwelling type, and maximum entropy 1.94 means indicates there are seven dwelling types present in that area with an equal proportion in each. 

\section{Results}
Below are our main findings based on our analysis using the entropy measure as well as the conventional correlation statistic. 

\textit{1. Based on the entropy calculation, the core of inner London Boroughs is very homogeneous (low entropy) with heterogeneity filtering out towards the peripheries (high entropy). The outer London Boroughs are dominated by diverse dwelling types (mid to high entropy values). When correlated, the locations of Airbnb establishments are negatively associated with dwelling type diversity and are positively associated with flats (purpose-built, conversions and flats in commercial buildings).}

At first glance, we might assume that central areas have a mix of dwelling types because the general conception is that cities are very diverse in the centre and homogeneous on their edge. But in the case of dwelling types, although the housing typology in the city centre might be more diverse, the buildings are occupied in a quite homogeneous way. The area inside the white boundary in Figure \ref{fig:entropy} defines the inner boroughs, and the core of that area spanning east has an extremely low entropy value (0.05 to 0.62) indicating that the area is dominated by a single dwelling type. The areas with the lowest entropy are the City of London, Hackney, Tower Hamlets and part of Kensington, and Lambeth, as well as a fraction of Westminster. Looking back historically, inner and east London have experienced heavy de-industrialisation in the past and other land use change which has allowed space for newer development. These are the areas where houses have been built after the 1970s. Moreover, 31\% of the total housing built in 2000 or later is located in Tower Hamlets \citep{londonhousing}. These new developments are mostly purpose-built blocks of flats to cater for singles, couples and small families. Aside from the core area, London's dwelling types are quite diverse generally with mid to high diversity as shown by the entropy statistics.

\begin{figure}[ht!]
        \centering
            \includegraphics[width = 0.50\textwidth]{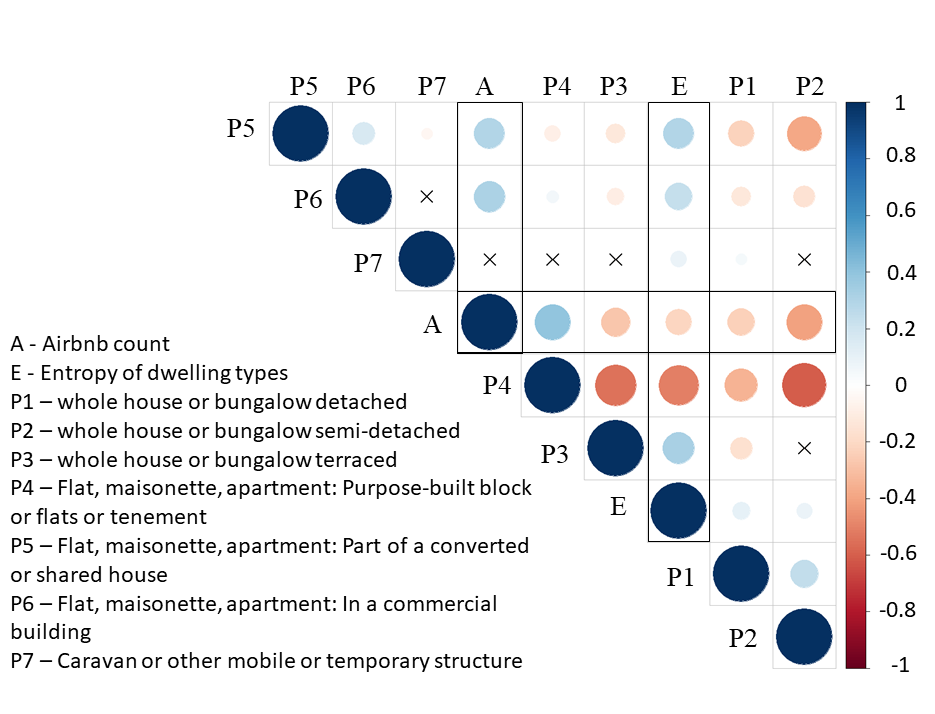}
    \caption{Correlations of Airbnb and dwelling types metrics (\textit{p}-value $<$ 0.01)}
    \label{fig:pearson}
\end{figure}

Furthermore, we use correlation to measure the strength (decreasing or increasing) of a linear relationship between variables \citep{lee1988thirteen}. The coefficient values can range from -1.00 and +1.00 and Figure \ref{fig:pearson} shows the correlogram depicting the correlation coefficients between Airbnb and the various dwelling types as well as showing their entropy measures in London. The matrix is used to investigate any simultaneous dependence between these multiple variables. The blue circles indicate positive and the red circles indicate negative associations. The colour intensity and the size of the circle are proportional to the correlation coefficients according to the colour bar legend on the right-hand side. Any non-significant results with a \textit{p}-value higher than 0.01 are deleted from the analysis. 

The analysis shows that the number of Airbnb listings is negatively correlated with the entropy of the dwelling type in terms of the correlation coefficients (\textit{r} of -0.22), indicating that Airbnb is associated with more homogeneous dwellings. Examining further, Airbnb listings have a positive association with specific dwelling types such as purpose-built flats/apartments, conversions, and flats in commercial buildings (\textit{r} of +0.40, +0.31 and +0.30 respectively). In contrast, the number of listings has a negative association with semi-detached, terraced and detached whole houses (with \textit{r} of -0.40, -0.27 and -0.24 respectively) as shown in Figure \ref{fig:pearson}. We can conclude from our analysis that the concentration of Airbnb tends to be in areas with a higher proportion of flats and apartments, a lower proportion of whole houses and in areas where there is lower dwelling type diversity (lower entropy).

\begin{figure}
    \centering
    \begin{subfigure}{0.40\textwidth}
            \includegraphics[width = 1\textwidth]{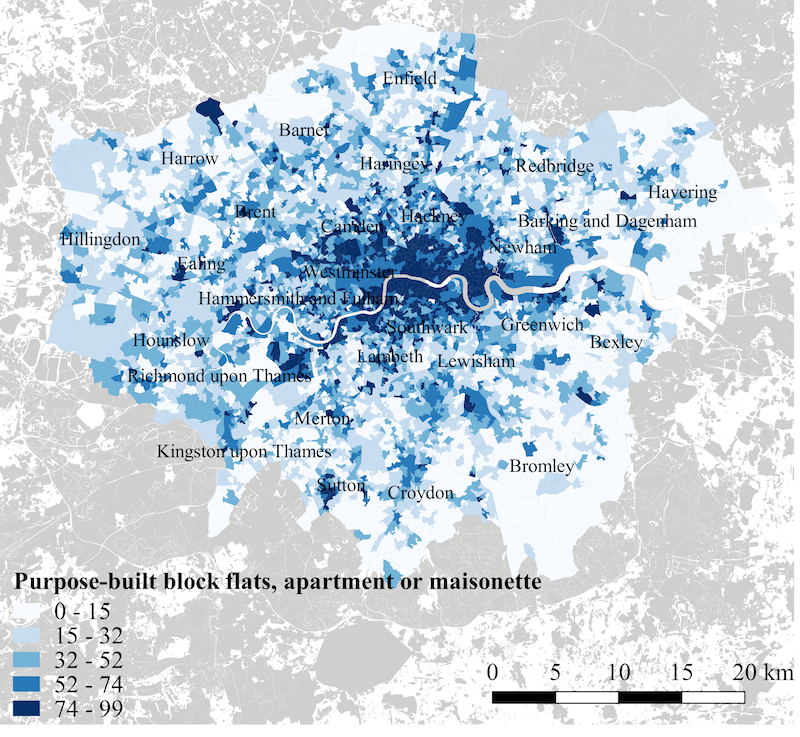}
            \caption{Percentage of purpose-built blocks flats, \textit{r} = 0.40}
            \label{fig:purposebuilt}
    \end{subfigure}
     \begin{subfigure}{0.40\textwidth}
        \centering
            \includegraphics[width = 1\textwidth]{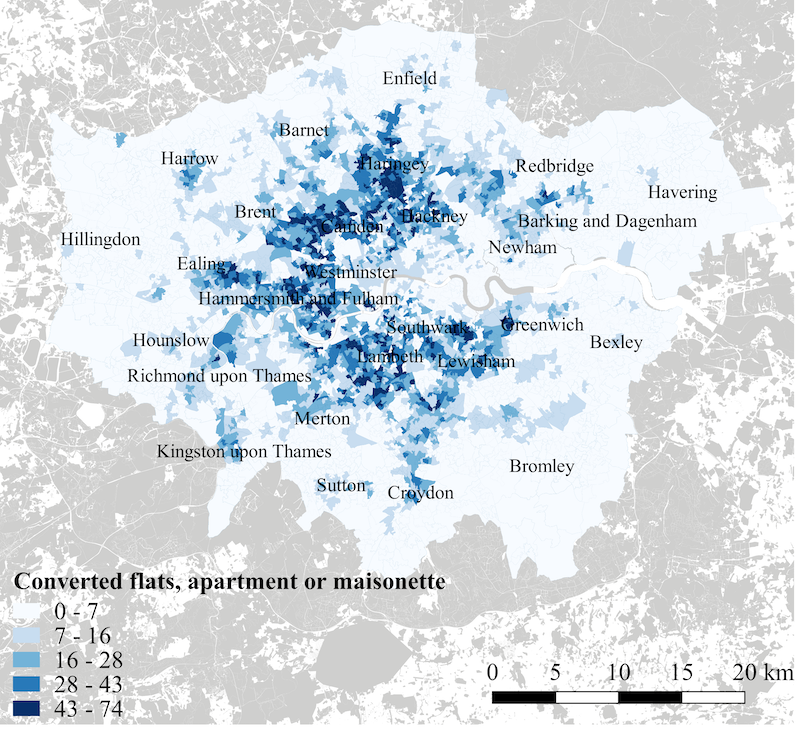}
            \caption{Percentage of conversion flats, \textit{r} = 0.31}
            \label{fig:converted}
    \end{subfigure}
    \begin{subfigure}{0.40\textwidth}
        \centering
            \includegraphics[width = 1\textwidth]{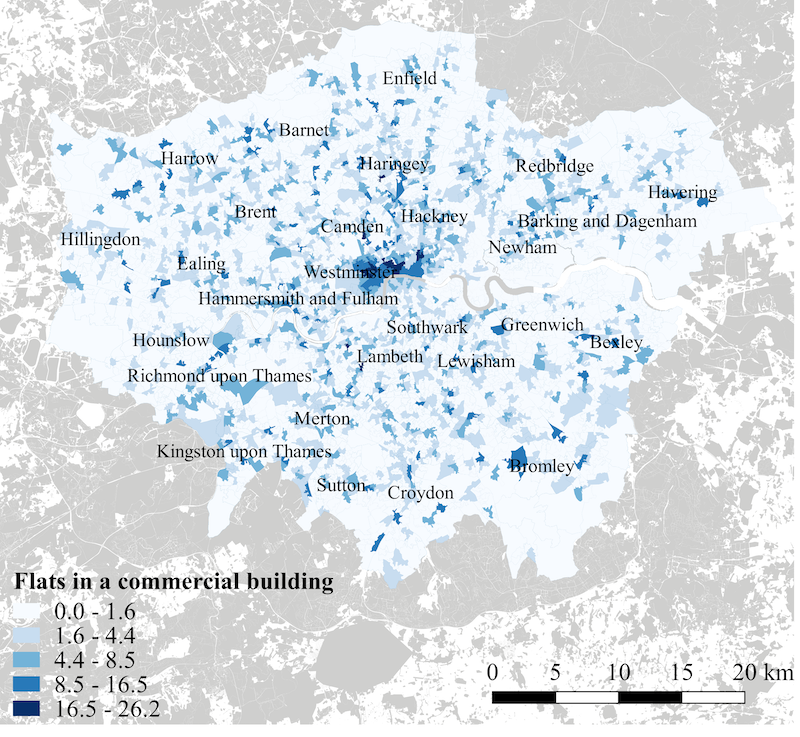}
            \caption{Percentage of flats in commercial buildings, \textit{r} = 0.30}
            \label{fig:commercial}
    \end{subfigure}
    
    \caption{The distribution of flats, apartments and maisonettes in London (the sub-captions showing the correlation coefficients \textit{r} between Airbnb and the specific dwelling type)}
    \label{fig:flats}
\end{figure}

Considering that the distribution of Airbnb shows a positive correlation with flats, apartments and maisonettes, Figure \ref{fig:flats} further displays the distributions of these specific dwelling types. Figure \ref{fig:purposebuilt} shows the distribution of purpose-built blocks of flats (flats that have been constructed originally as flats as opposed to conversions from other functions, for example). The main characteristics of purpose-built flats are that they have regular shapes, most are newly built, and they are generally newer than converted flats. They can be either low rise (usually less than six storeys) or high rise (more than six storeys). Airbnb has the highest correlation with purpose-built flat dwellings compared to other types of dwellings where there is a positive linear relationship between the two indicated by r of +0.40. Although spatially distributed all over London, the highest proportion of purpose-built flats is 99\%, mostly in Tower Hamlets, the City of London and Hackney, and there is a high proportion of flats generally distributed through Central London towards East London with some increase in proportions in Greenwich. These overlap with the areas with the highest concentrations of Airbnb.

The Airbnb count also shows a positive correlation with conversions. The demographic structure in London was changing from the 1980s to 1990s with a rise in one and two-person households, and this led to many whole house properties being converted into flats \citep{hamnett1973improvement,hamnett2004unequal}. For example, many two or three storey terrace houses were broken up into several flats in that time period. Based on an LRC (London Research Centre) report, all areas across inner London and beyond that have potential for profitable conversion have undergone changes particularly those in prime areas. It is estimated there was a total of 120,700 units converted in a short period of time from 1980 to 1989 \citep{hamnett2004unequal}. Moreover, former industrial properties and offices especially those in close proximity to the Thames and inner-city London were converted into flats as well, further increasing London's housing supply. From Figure \ref{fig:converted} we can see that the areas with high proportions of converted flats are located in West London, North London and South London near the river. In the 2000s, London experienced a substantial increase in its housing stock, and this was not due to the construction of new houses, but rather from a combination of conversions of existing houses to flats and buildings changing from industrial/commercial to residential \citep{londonhousing}. This might be a London phenomenon as London is very different from the rest of the UK in that  over half of London housing comprises flats (see again Table \ref{tab:dwellingtype} for more details) compared to less than 20\% in the rest of the country \citep{londonhousing}.

Figure \ref{fig:commercial} shows the distribution of flats in commercial buildings. The proportion of this type of dwelling is generally lower compared to other type of flats (the highest is 26\% of the overall dwellings, located in an LSOA in Westminster), and they are dispersed all around London. Overall, the correlation coefficient shows the positive linear correlation of Airbnb establishments with flat dwelling type as shown in Figure \ref{fig:flats}, even though not all those dwellings are located centrally.

\textit{2. Airbnb is located in areas with a high proportion of privately rented sector.}

\begin{figure}[ht!]
        \centering
            \includegraphics[width = 0.50\textwidth]{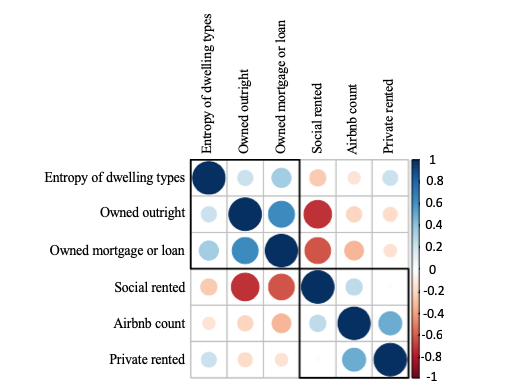}
    \caption{Correlation of Airbnb, entropy of dwelling types and the type of home ownership ($p$-value < 0.01)}
    \label{fig:lsoa}
\end{figure}

In their previous work, \cite{quattrone2016benefits} found that Airbnb offerings correlate with the socio-economic attributes of their geographical locations. These are the areas that have high proportions of young population that are ethnically mixed and economically active with a low proportion of owned properties and good accessibility to the city centre \citep{quattrone2016benefits}. Expanding on their previous work, we look into greater details the relationship between the entropy of dwelling types, Airbnb, and type of home ownership in London as shown in Figure \ref{fig:lsoa}. The analysis shows that areas with a higher diversity of dwelling types (mid to high entropy) can be associated with owned properties (both outright (\textit{r} = +0.22) and owned with mortgages/loans (\textit{r} = +0.34)). In contrast, Figure \ref{fig:lsoa} shows that the Airbnb count has a positive correlation (\textit{r} = +0.50) with areas that have a higher proportion of private rented properties and a negative linear relationship with owner-occupied dwellings (\textit{r} owned outright = -0.22 and \textit{r} owned mortgage =-0.33). In London, private renting was once the largest tenure until the 1980s, when there was a dramatic transformation to ownership where almost 40\% of households in inner London either own or are buying their own homes \citep{hamnett2004unequal}. Recently, the number of private rentals has been rising rapidly to 26\%, making it the second largest tenure, while owner occupancy has fallen in commensurate terms \citep{londonhousing}. The median price of this private rented sector varies hugely between various boroughs, where higher rents tend to be located in inner London boroughs (areas with large rented sector) \citep{londonhousing}. There are thus very distinct characteristics for areas with high entropy and those with high numbers of Airbnb establishments.

There has been a rise in lawsuits associated with illegal listings where Airbnb establishments are being advertised by renters without the acknowledgement of the property owners \citep{oliviarudgard2016,donaferguson2016}. Considering that our analysis finds that Airbnb can be associated with a higher proportion of privately rented properties, this is not so surprising as might be seen at first sight. Generally, a tenancy agreement puts a clause in to prevent subletting without the owner's consent. Only the outright owners have the ability to rent out their property freely, as many mortgage providers also do not allow subletting without their consent, and would add a surcharge to the mortgage interest rate \citep{ameliamurray2017}. Our second findings call for a more detailed study of how many Airbnb establishments are operating illegally, bypassing the basic tenancy contract. The implications can be threefold: firstly property owners face the risks of damaged properties due to inappropriate conversion from long to short term rentals, and secondly there are risks of eviction for renters who illegally rent out the properties as well as the guests who stay in the illegal Airbnb establishments.

Thirdly, another implication is that many Londoners have been priced out and have been moving to the commuter belt areas due to the annual increase of rent in London \citep{isabellefraser2017}. The problem of affordability is even worse in the prime London LSOAs, where Airbnb are heavily concentrated, as in the areas with mid to high rental prices. A report shows that in August 2018, there was a 24\% drop in available properties to rent in Greater London compared to the previous year; thus an increase in rent in the future is expected \citep{marcdasilva2018}. Airbnb is likely to have exacerbated this problem.

\section{Conclusion}

The concept of peer-to-peer renting through the emerging online platform economy, such as that developed by Airbnb, has various implications for the long term housing supply. This problem is even more urgent in a city where the housing market is overheated and complex such as in London, as it has been very difficult, especially for the working class and low income groups to compete in the housing market. The conversion from long-term housing to short-term rentals facilitated by Airbnb further worsen this problem. This paper has revealed this gap regarding how Airbnb is changing the structure of cities, especially in terms of housing supply, using London as a case study.  

Based on our diversity measure of dwelling types using the concept of entropy, London's core area is quite homogeneous (especially in its inner city areas and East London) whereas elsewhere there is mid to high diversity. Past research tends to associate diversity of the city centre in terms of land use but this is very different to the picture we have presented in this paper. We focus on the diversity of dwelling types and find that the dwellings in London's city centre are being used in a very homogeneous way. This finding can be linked to the history of housing in London with its concentration of purpose-built flats dominating the central to eastern areas since 1980s onward, resulting in low diversity (low entropy) in those areas. We also find that the diversity of dwelling types is negatively correlated with Airbnb locations, as Airbnb tends to locate in areas that are dominated by flats (purpose-built, conversions and flats in commercial buildings) as can be seen by the positive correlation between Airbnb locations and flats.

Areas with a high proportion of flats such as Hackney, Tower Hamlets and Westminster, are those that are more prone to be converted to short-term rentals compared to areas with predominantly whole house dwellings. Looking at our analysis on Airbnb density, more than 20\% of the housing supply has been advertised on the Airbnb platform in an LSOA in Hackney. Although the overall proportion of Airbnb only accounts for 1.4\% of the overall housing supply, but because Airbnb establishments are only concentrated in certain areas, the effect of Airbnb can be much greater in some LSOAs in London.  

We also find that the number of Airbnb establishments also correlates with a high proportion of privately rented properties (\textit{r} = 0.50). More than 26\% of London's population are renting their properties, and it is also possible that Airbnb is contributing to the drop in the number of rental homes by detracting from the supply of long-term housing. The implication of this finding is also related to the legalities of those listings, as illegal listings add unintended risks to all homeowners, renters and guests. 

We consider there are several limitations to our analysis in this paper. As the study has only considered the London case, it is not possible to draw a more general conclusion as Airbnb case is city-specific. Therefore, it would be interesting to see if this is also the case in other major cities with complex housing problems such as Paris, New York and Berlin. Also, this study specifically analyses the diversity of dwelling types, thus disregarding other types of diversity such as that associated with land use. Incorporating different metrics to analyse the geographical patterns of Airbnb at different scales would yield interesting results which would hopefully complement this analysis and deepen our understanding of this phenomenon further. Another important limitation is the challenge as to how we might quantify how many people are displaced due to Airbnb conversions.

Although to some extent we have already started to touch upon some of the effects of Airbnb on housing in general, following the work of \cite{quattrone2016benefits, wachsmuth2017airbnb, horn2017home, lee2016airbnb, jefferson2014airbnb, schafer2016misuse}, more perspectives from different cities are needed to enrich our knowledge of Airbnb and housing. Our own topics for future studies include identifying areas that are vulnerable to Airbnb conversion, analysing how Airbnb affects the prices for long-term housing provision, and formulating proper regulatory controls for Airbnb establishments. It is important to conduct such critical analyses in order to overcome the challenges that come with rapid developments of new phenomena such as short-term rentals without stifling this kind of innovation by imposing mandatory regulations that are not thought through. Platforms such as Airbnb are a key representation of the wider 'sharing' economy which extends from mobility issues to many kinds of retailing and capital investment such as housing. We believe we have made a start on this problem in terms of the focus in this paper.

\bibliographystyle{apalike}

%referencing the bib file for bibliography

\end{document}